**Cascade of electronic transitions in magic-angle twisted bilayer graphene**

Dillon Wong[1,*], Kevin P. Nuckolls[1,*], Myungchul Oh[1,*], Biao Lian[2,*], Yonglong Xie[1,†,§], Sangjun Jeon[1,#], Kenji Watanabe[3], Takashi Taniguchi[3], B. Andrei Bernevig[2], and Ali Yazdani[1,‡]

[1]*Joseph Henry Laboratories & Department of Physics, Princeton University, Princeton, NJ 08544, USA*
[2]*Princeton Center for Theoretical Science, Princeton University, Princeton, New Jersey 08544, USA*
[3]*National Institute for Material Science, 1-1 Namiki, Tsukuba 305-0044, Japan*

\* These authors contributed equally to this work.
† Present Address: Department of Physics, Harvard University, Cambridge, MA 02138, USA
§ Present Address: Department of Physics, Massachusetts Institute of Technology, Cambridge, Massachusetts 02139, USA
# Present Address: Department of Physics, Chung-Ang University, Seoul 06974, Republic of Korea
‡ email: yazdani@princeton.edu

**Magic-angle twisted bilayer graphene (MATBG) exhibits a rich variety of electronic states, including correlated insulators[1–3], superconductors[2–4], and topological phases[3,5,6]. Understanding the microscopic mechanisms responsible for these phases requires determining the interplay between electron-electron interactions and quantum degeneracy due to spin and valley degrees of freedom. Signatures of strong electron-electron correlations have been observed at partial fillings of the flat electronic bands in recent spectroscopic measurements[7–10]. Transport experiments have shown changes in the Landau level degeneracy at fillings corresponding to an integer number of electrons per moiré unit cell[2–4]. However, the interplay between interaction effects and the degeneracy of the system is currently unclear. Using high-resolution scanning tunneling microscopy (STM), we observed a cascade of transitions in the spectroscopic properties of MATBG as a function of electron filling. We find distinct changes in the chemical potential and a rearrangement of the low-energy excitations at each integer filling of the moiré flat bands. These**

**spectroscopic features are a direct consequence of Coulomb interactions, which split the degenerate flat bands into Hubbard sub-bands. We find these interactions, the strength of which we can extract experimentally, to be surprisingly sensitive to the presence of a perpendicular magnetic field, which strongly modifies the spectroscopic transitions. The cascade of transitions we report here characterizes the correlated high-temperature parent phase[11,12] from which various insulating and superconducting ground-state phases emerge at low temperatures in MATBG.**

Stacks of two-dimensional (2D) monolayers represent a new platform for the realization of correlated and topological electronic states. By twisting 2D crystals relative to each other, it is possible to form a moiré superlattice with electronic bandwidth comparable in energy to that of Coulomb interactions, thereby creating the conditions for the formation of correlated phases[1,2,4,13–17]. Recent discoveries of various electronic phases in MATBG have occurred precisely in this regime, where band structure calculations show the formation of two nearly flat bands near charge neutrality[18–22], while estimates for the Coulomb interaction strength yield an energy comparable to or larger than the expected flat bands' bandwidth[23,24]. Consistent with this picture, previous STM measurements have shown signatures of many-body correlations in MATBG[7–10]. Spectroscopy at densities where the flat bands are fully occupied or empty show sharp van Hove singularities (vHs) associated with nearly flat conduction and valence bands, while spectra of partially occupied flat bands show dramatic broadening of both bands, indicating the importance of interaction-induced charge fluctuations[7]. Here we discover a sequence of spectroscopic transitions at every integer filling of both flat bands. An analysis of our spectroscopic data based on a Hubbard-model picture of its electronic states provides a direct measurement of the Coulomb repulsion strength $U$ and uncovers signatures of sharp transitions in the chemical potential at every integer filling. Our results demonstrate the influence of spin and valley degeneracies on the electronic properties of this strongly interacting system.

A schematic of our experimental setup is shown in Fig. 1a. MATBG samples were prepared on boron nitride (h-BN) using the "tear-and-stack" technique[25] (see Methods) and were electrically contacted via Ti/Au electrodes on $SiO_2$/Si. Our devices were studied in a homebuilt STM[26], biased at $V_b$ relative to a tungsten tip, and gated with a voltage $V_g$ applied to Si. An STM topograph of a typical device (Fig. 1b) displays three coexisting lattices: the graphene atomic lattice, a moiré pattern of size $\lambda_{g\text{-BN}} \approx 1.76$ nm (8°) due to misalignment of the graphene and h-BN, and a moiré pattern of size $\lambda_{g\text{-}g} \approx 13.3$ nm (1.06°, 0.1% interlayer relative strain) due to misalignment of the top and bottom graphene layers. The bright regions in the image correspond to AA stacking regions of MATBG[27–29], where the differential conductance (d$I$/d$V$) shows the vHs of the flat bands appearing as two peaks whose energy locations and widths depend on the carrier density (Extended Data Fig. 1).

A cascade of changes in the electronic properties of MATBG are observed in spectra measured at the center of each AA site as a function of $V_g$ (Fig. 1c), which tunes the electron filling $\nu = \pm n/n_0$, where $\nu = \pm 4$ corresponds to fully occupied/empty fourfold-degenerate (spin/valley) flat bands. As seen previously[7–9], the vHs of the two flat bands appear as two sharp, parallel lines in d$I$/d$V$($V_b$, $V_g$) when both flat bands are fully occupied ($V_g > 40$ V) or empty ($V_g < -40$ V). In this region, d$I$/d$V$ is consistent with non-interacting models of the band structure of MATBG[18–21]. At $\nu = \pm 4$, when the chemical potential is tuned across the gap from either of the remote bands into either of the flat bands, the spectra show immediate broadening of the flat bands, which is a consequence of strong electronic interactions[7]. This broadening is accompanied by a sequence of previously unresolved finer features around the Fermi energy ($E_F$) (Fig. 1c).

Division of d$I$/d$V$ by $I/V_b$, typically used when there are large variations of the tunneling current, more clearly displays these finer features (Fig. 1d). The cascade of features we observe around $E_F$ repeats in four evenly spaced intervals between $\nu = +4$ and the charge neutrality point (CNP) at $\nu = 0$, and in four more intervals before the chemical potential is tuned below the flat bands and into the valence remote band at $\nu = -4$. The 14 features seen around $E_F$, schematically sketched in Fig. 1e, show systematic variations with $V_b$ as a function of filling. Each unfilled excitation ($V_b > 0$ V) appears first at $E_F$ and disperses outwards with decreasing $\nu$ to energies of about 23 meV before disappearing. The features near the CNP disperse to higher energies, pushed further by the energy between the flat bands. In contrast, each filled excitation ($V_b < 0$ V) is first detected at negative bias and moves closer to $E_F$ with decreasing $\nu$ (Figs. 1c-f). Extended Data Fig. 2 shows data acquired on AA sites in twisted regions within 0.1° of the magic angle containing features with similar behavior.

A starting point for understanding the sequence of features in our measurements is to consider a model consisting of two perfectly flat bands separated by energy $2E_0$, each with four internal degrees of freedom ("flavors") due to spin and valley degeneracies, and an on-site Hubbard interaction $U$ describing the Coulomb repulsion between electrons (Fig. 2a). As described below, our data suggests that the bands are not perfectly flat and that there is a small but finite density of states (DOS) between the flat bands (potentially from Dirac nodes expected in MATBG's band structure). However, this simple Hubbard-model picture in the limit of vanishing hopping provides a natural way to understand our observations. In Fig. 2a, there are sudden changes in the spectral weight corresponding to adding and removing an electron at integer $\nu$. There are also jumps in the chemical potential at integer $\nu$ (Fig. 2d) equal to $U$ (except at the CNP, where the chemical potential jumps by $2E_0 + U$).

The features in the calculated spectral weight (Fig. 2a) arise from one-particle excitations from nine types of minimum-energy states schematically depicted in Fig. 2b. The red (blue) ovals represent the conduction (valence) flat band, each colored circle represents an electron of a specific flavor, and each arrow represents a possible excitation from this minimum-energy state and its corresponding energetic cost. Together, these excitations give rise to the "Hubbard sub-bands" drawn in Fig. 2c whose spectroscopic degeneracy is given by the number of ways an electron can be added to or removed from a minimum-energy state. The on-site repulsion $U$ separates each fourfold-degenerate flat band into Hubbard sub-bands, even in the absence of symmetry-breaking effects. Within this picture, the features corresponding to electron addition excitations peel away from $E_F$ between integer fillings, reaching a maximum energy of $U$ before losing intensity. The spectral weight beyond energy $U$ is generically from one-particle excitations of higher-energy on-site electron configurations and is exponentially suppressed (see Methods). Near the CNP, the features corresponding to electron addition or removal also include the energy separation between the flat bands, causing them to extend to $\pm(2E_0 + U)$. This enhanced flat band separation is consistent with the features seen in Figs. 1c-d near $V_g = 0$ V, which extend to higher biases than those seen at other filling values.

There are a few key differences between this perfectly flat band Hubbard model and our observations. First, from our measurements, we expect the two flat bands to have finite bandwidths with a small but finite DOS between the bands. This would cause the cascading features to disperse at intermediate fillings instead of being pinned at $E_F$ and suddenly jumping at integer $\nu$, which is more consistent with the experimental observations. Exact diagonalization with a finite hopping term clearly demonstrates this behavior (Extended Data Fig. 3). Second, the electric field caused by the bias voltage and tip-sample work function difference causes local band bending, slightly shifting some of the features in Fig. 1c from the idealized picture in Fig. 2a[7].

From the correspondence drawn between the features in our data and the excitation spectrum expected from the Hubbard model, we extract a measure of the on-site Coulomb repulsion strength $U = 23 \pm 5$ meV, consistent with previous estimates[7]. This value of $U$ is much larger than non-interacting estimates for the bandwidth of a flat band and the energy separation between two flat bands[18–21], both of which we have also experimentally resolved in the non-interacting regime ($V_g > 40$ V in Fig. 1c). This relation inverts for twist angles larger and smaller than the magic angle, where the bandwidth becomes comparable to or larger than the interaction energy. In this regime, for 0.85° and 1.21° twisted bilayer graphene, we find no evidence of the cascades despite the presence of strong $E_F$ pinning behavior (Extended Data Fig. 4). This confirms the lack of well-defined Hubbard sub-bands when $U$ is not the largest energy scale.

The flavor-induced reorganization of electronic states in MATBG can also be seen in the gate dependence of the spectroscopic behavior of the remote bands nearest to the flat bands, from which we can extract information regarding changes in the chemical potential as a function of carrier density. To probe the remote bands, we measure d$I$/d$V$ at the center of AB/BA regions (Fig. 3a; 1.05°, 0.2% strain), where the flat bands' contribution to the spectral weight no longer dominates the measurements. Spectra acquired at these locations provide information that is more sensitive to the delocalized states of MATBG and their density-dependent interactions with the flat bands. The edge of the conduction remote band, with onset energy $E_{R+}(\nu)$ (37.5 meV at $\nu = +4$), and the valence remote band, with onset energy $E_{R-}(\nu)$ (-79.5 meV at $\nu = +4$), show a periodic modulation as a function of density that is in-phase with integer fillings of the flat bands (Figs. 3a, c). Such behavior is most clearly resolved in the conduction remote band above the CNP.

To understand the gate-dependent periodicity of the remote band onsets, we consider a simple mean-field interaction model $E_{R+(-)}(\nu) = \Delta_{R+(-)} + \nu U' - \mu(\nu)$, where $\Delta_{R+(-)}$ is the energy difference between the flat bands and the conduction (valence) remote band in the absence of any interactions, $\nu U'$ is a mean-field term describing the interactions between electrons in the flat bands and an electron in the remote band (the mean-field approximation is justified because the remote bands are delocalized), and $\mu(\nu)$ is the chemical potential at filling $\nu$. In the absence of interband interactions, $E_{R+(-)}(\nu)$ directly measures $\mu(\nu)$, which we plot in Fig. 3c. To include a mean-field treatment of the interband interactions, we use the full form for $E_{R+(-)}(\nu)$ to extract the interaction strength $U'$ from the data in Fig. 3a and determine the behavior of $\mu(\nu)$ (see Methods). The results (Fig. 3d) illustrate that while $\mu(\nu)$ does not jump at integer $\nu$ as we expect for our simple Hubbard model (Fig. 2d; Extended Data Fig. 3b), $\mu(\nu)$ shows distinct cusps at every integer filling. The difference lies in the fact that, at integer fillings, our simple Hubbard model in the limit of $U$ much greater than the bandwidth $W$ is a Mott insulator, which has a charge gap of size $U-W$ at $E_F$. In contrast, our measurements at these high temperatures ($T \approx 6$ K) do not yet show any hard gaps at $E_F$, indicating a finite DOS between the flat bands. This finite DOS could be associated with, for example, the tails of the flat bands or a high-velocity Dirac point between these bands. The cusps in $\mu(\nu)$ at integer $\nu$ also coincide with dips in the DOS at $E_F$ that can be seen in the plot of $dI/dV(V_b = 0$ V, $V_g)$ in Fig. 3e. Moreover, each cusp forms at the junction of two asymmetric slopes of $\mu(\nu)$, indicating differences in the DOS between the top and bottom of sequential Hubbard sub-bands (schematically depicted in Fig. 2c). The cusps in $\mu(\nu)$ imply a sudden drop in the compressibility and the cyclotron mass when increasing (decreasing) the carrier density across a positive (negative) integer value of $\nu$. This observation may explain the unusual asymmetry in the Landau fans seen near non-zero integer fillings in transport studies[2–4].

We have thus far only invoked local, on-site considerations of the Coulomb interactions between electrons, motivated by the experimental observation that the spectral weight of the flat bands is mostly spatially localized to the AA sites. Such a local picture is adequate for understanding most of our findings, yet fails to explain the behavior of our system in the presence of a large magnetic field. Surprisingly, we find that an out-of-plane magnetic field of 9 T (at $T \approx 6$ K) strongly modifies the observed zero-field cascades and suppresses both the cusps in $\mu(\nu)$ at integer $\nu$ and the dips in the DOS at $E_F$ (Fig. 3b, e; see Extended Data Fig. 5 for AA data at 9 T). A large response to low magnetic fields has been predicted to arise from coupling to the magnetic moment of moiré-scale orbital current loops that form due to valley polarization[30–32]; however, the relationship between our measurements and these proposed ground states is unclear. This remarkable sensitivity to a perpendicular magnetic field, in which changes in degeneracy have been detected in lower-temperature transport studies[2–4], cannot be understood with our local picture. The experimental data suggest that application of a perpendicular magnetic field of this magnitude may actually delocalize some electrons within the flat bands from the confinement of the moiré superlattice structure, which would result in an apparent suppression of the effective local interaction strength $U$ that drives the cascade of transitions and remote band oscillations that we have uncovered at zero field.

Further measurements are required to fully establish the connection between our high-temperature observations of the restructuring of low-energy excitations, the cusps in $\mu(\nu)$, and the suppression of DOS at $E_F$ at each integer filling to the formation of insulating and magnetic phases reported in transport studies at lower temperatures[1–4]. Our experiments show that even at high temperatures, interactions separate the flat bands of MATBG into several Hubbard sub-bands. This could further drive the formation of long-range flavor-polarized insulating or magnetic ground states at low temperatures, as several theoretical studies suggest[23,33]. A tendency towards flavor polarization near integer fillings might favor unconventional superconductivity if such polarization is also present between integer fillings. However, the presence of strong interactions may also cause the spontaneous breaking of other symmetries to create insulating phases different from those with purely spin/valley flavor polarization[21]. More broadly, we speculate that similar cascades of transitions may occur in other moiré flat band systems, where local electronic interaction effects are dominant and internal quantum degrees of freedom are present.

**Acknowledgements**

We thank P. Jarillo-Herrero, E. Berg, A. Stern, A.H. MacDonald, B. Jäck, X. Liu, C.-L. Chiu, N.P. Ong, S. Wu, S. Todadri, and S. Kahn for useful discussions. This work was primarily supported by the Gordon and Betty Moore Foundation as part of the EPiQS initiative (GBMF4530) and DOE-BES grant DE-FG02-07ER46419. Other support for the experimental work was provided by NSF-MRSEC through the Princeton Center for Complex Materials DMR-1420541, NSF-DMR-1608848, NSF-DMR-1904442, ExxonMobil through the Andlinger Center for Energy and the Environment at Princeton, and the Princeton Catalysis Initiative. K.W. and T.T. acknowledge support from the Elemental Strategy Initiative conducted by the MEXT, Japan, A3 Foresight by JSPS and the CREST (JPMJCR15F3), JST. B.L. acknowledges support from the Princeton Center for Theoretical Science at Princeton University. B.A.B. acknowledges support from the Department of Energy DE-SC0016239, Simons Investigator Award, the Packard Foundation, the Schmidt Fund for Innovative Research, NSF EAGER grant DMR-1643312, and NSF-MRSEC DMR-1420541. AY acknowledges the hospitality of the Trinity College and Cavendish Laboratory in Cambridge, UK during the preparation of this manuscript, which was also funded in part by a QuantEmX grant from the Institute for Complex Adaptive Matter and the Gordon and Betty Moore Foundation (GBMF5305).

**Author Contributions**

D.W., K.P.N., M.O., and A.Y. designed the research strategy, carried out STM/STS measurements, and performed the data analysis. D.W., M.O., and K.P.N fabricated samples. S.J., D.W., K.P.N., M.O., and A.Y. constructed the STM. K.W. and T.T. synthesized the h-BN crystals. B.L. and B.A.B. performed the theoretical calculations. All authors discussed the results and contributed to the writing of the manuscript.

**Figure Captions:**

**Figure 1 | Cascade of transitions in spectroscopy on an AA site. a**, Schematic of the experimental setup. **b,** STM topographic image of 1.06° MATBG ($V_b$ = -80 mV, $I$ = 300 pA). **c,** $dI/dV(V_b, V_g)$ measured at the center of an AA site (Initial tunneling parameters: $V_b$ = -100 mV, $I$ = 1 nA, 4.121 kHz sinusoidal modulation of $V_{rms}$ = 1 mV). **d**, Same as **c** but divided by $I/V_b$. **e,** Cartoon schematic of features seen in **c** and **d**. **f**, $dI/dV$ line cuts from $V_g$ = 31 V to 13 V (left) and $V_g$ = -4 V to -23 V (right). Triangles mark peaks corresponding to the features schematically depicted in **e**. Sequential curves are vertically offset by 2.5 nS (left) and 4 nS (right) for clarity.

**Figure 2 | Theoretical model with on-site Hubbard repulsion *U*. a,** Calculated spectral weight for a two-band Hubbard model with four flavors per band in the limit of vanishing hopping. The

on-site interaction strength is $U = 23$ meV, the energy difference between the flat bands is $2E_0 = 16$ meV, and the temperature is $k_B T = 0.5$ meV. **b,** The features in **a** correspond to the on-site one-particle excitations schematically depicted here. The colored circles correspond to electrons of different flavors, and the arrows represent different possible excitations. Each excitation has two possible energies depending on the value of the chemical potential (determined by the sign of δ). **c,** Schematic for the spectroscopic degeneracy (SD) versus energy ($E$) for various integer values of $\nu$. The SD is determined by the number of excitations (arrows) in **b**, and the colors indicate the types of excitations depicted in **b**. **d,** The calculated chemical potential as a function of filling jumps by $U$ at integer $\nu$, except at the CNP where it jumps by $2E_0 + U$.

**Figure 3 | Chemical potential observed through remote band oscillations. a,** $dI/dV(V_b, V_g)$ measured at the center of an AB site at zero magnetic field (Initial tunneling parameters: $V_b = -100$ mV, $I = 1$ nA, 4.121 kHz sinusoidal modulation of $V_{rms} = 0.6$ mV). **b,** $dI/dV(V_b, V_g)$ measured at the center of an AB site at out-of-plane magnetic field $B_\perp = 9$ T (Initial tunneling parameters: $V_b = -100$ mV, $I = 0.6$ nA, 4.121 kHz sinusoidal modulation of $V_{rms} = 0.6$ mV). **c,** Onset energy of the conduction (pink) and valence (cyan) remote bands (RB) as a function of $V_g$ extracted from data in **a** (solid lines) and in **b** (dotted lines). Curves are continuous at $\nu = 0$, but curves on the left and right panels were offset by different arbitrary energies for clarity. In the absence of interband interactions ($U' = 0$), the remote band onsets directly measure the chemical potential. **d,** Chemical potential calculated from data in **c** using the mean-field model discussed in the text (with $U' = 15$ meV). Cusp-like features, which signify discontinuities in the electronic compressibility, are seen at each non-zero integer filling. As in **c**, curves are continuous at $\nu = 0$ but are offset by arbitrary energies for clarity. **e,** $dI/dV(V_b = 0$ V, $V_g)$ from the data in **a** ($B_\perp = 0$ T;

solid line) and **b** ($B_\perp = 9$ T; dotted line). At $B_\perp = 0$ T, dips in conductance are visible at each integer filling, denoted by the set of nine evenly spaced shaded bars.

**Methods**

**STM measurements.** All STM measurements were performed on a homebuilt, ultra-high vacuum (UHV) STM[26] near $T = 6$ K using a tungsten tip. The tip was prepared on a Cu(111) single crystal and calibrated against the crystal's characteristic Shockley surface state (see Extended Data Fig. 6 for details about why this is necessary). Graphene was electrostatically gated through a voltage $V_g$ applied to a degenerately p-doped Si back-gate, while $V_b$ was applied to the sample relative to the tip. Scanning tunneling spectroscopy (STS) measurements were obtained through lock-in detection of the tunneling current induced by a small sinusoidal modulation voltage added to the sample bias. The STM tip was navigated to graphene using a capacitance-based technique[34].

**Sample preparation**. Devices were prepared using a "tear-and-stack" method[25]. The fabrication procedure for the device that yielded the data in Fig. 1 ("Device A") is described here (see Extended Data Fig. 7 for a schematic). Transparent tape was pressed onto a glass slide. Two additional pieces of tape were placed on the slide parallel to the original tape and covering its edges to produce a channel running along the length of the glass slide. Polyvinyl alcohol (PVA, 5% by weight dissolved in water and filtered through a 0.2-μm pore-size PTFE syringe filter) was dropped onto the tape/glass slide. A separate glass slide was slid over the first glass slide to smooth out the PVA to the thickness of the channel. The PVA was allowed to dry in air for 15 minutes before the underlying transparent tape support was lifted from the glass slide. A 2 mm x 2 mm polydimethylsiloxane (PDMS, Sylgard 184) film was then placed between the PVA/tape and the glass slide to form the completed "PVA handle". Using a transfer station, the PVA handle was aligned with h-BN (exfoliated on SiO$_2$), which was picked up by the PVA at 60 °C. The PVA handle was then aligned such that this h-BN flake contacted half of a monolayer graphene sheet (exfoliated on piranha-cleaned $SiO_2$ using a hot cleave method[35]). As the PVA handle was lifted, graphene adhered to h-BN where contacted and tore at the h-BN flake's edge. The remaining graphene, left on the $SiO_2$ wafer, was rotated by 1.3° with respect to the PVA handle, and the PVA handle was translated to pick up the rotated piece of graphene. Care was taken throughout this procedure to remove mechanical backlash in the transfer station's rotation stage, ensuring higher precision of this rotation angle. A polymethyl methacrylate (PMMA)/tape/PDMS/glass slide (the "PMMA handle") was prepared in a similar manner to the PVA handle, omitting the channel construction steps to produce a thinner film. Additionally, the PMMA handle was baked on a hot plate at 130 °C for 5 minutes. Using the transfer stage, the PVA handle was aligned and contacted to the PMMA handle. Water was injected between the two handles with a syringe, dissolving the PVA and transferring the graphene/h-BN stack to the PMMA handle. The PMMA handle was placed in three beakers of room-temperature water for 10 minutes each, further dissolving any remaining PVA residue. Using the transfer stage again, the graphene/h-BN heterostructure on the PMMA handle was pressed against $SiO_2$/Si with pre-patterned Au/Ti electrodes at 110° C. The graphene at the edge of the h-BN flake makes electrical contact to these pre-patterned electrodes. After this transfer, PMMA residue was dissolved by placing the completed device in dichloromethane (DCM) heated just below its boiling point for 25 minutes. The device was subsequently dipped in acetone, water, and isopropanol (IPA) for a few minutes. Finally, the device was annealed in UHV for 12 hours at 170 °C, followed by a 2 hour anneal at 400 °C. The fabrication of the device that yielded data for Fig. 3 ("Device B") is similar, except the unfiltered PVA concentration was increased to 25% by volume, the PMMA was eliminated from the procedure, only water was used as the final solvent, the graphene was rotated by 1° during the tear-and-stack, and miscellaneous modifications were made to various temperatures and wait times.

**Theoretical calculations.** The zero-hopping-limit spectral weight in Fig. 2a was calculated by considering two orbitals at single-particle energies $\pm E_0$ corresponding to the two flat bands of MATBG. Each orbital has a 4-fold spin and valley degeneracy. At zero hopping, each site is decoupled from the others. The single-site Hamiltonian is given by

$$H = E_0(n_2 - n_1) + \frac{U}{2}n(n-1)$$

where $0 \leq n_\alpha \leq 4$ is the number of electrons in orbital $\alpha$ ($\alpha = 1, 2$), and $n = n_1 + n_2$ is the total number of electrons on an AA site. The spectral weight at temperature $T$ and chemical potential $\mu$ is given by

$$A(\omega) = Z^{-1} \sum_{l,m} \delta(\omega - \epsilon_l + \epsilon_m)\left(e^{-\beta\epsilon_m} + e^{-\beta\epsilon_l}\right)\left|\langle l|c^\dagger_{\alpha,f}|m\rangle\right|^2$$

where $Z = \text{tr}\, e^{-\beta(H-\mu n)}$ is the partition function ($\beta = 1/k_B T$), $\epsilon_m$ and $|m\rangle$ are the energy and wavefunction of the m-th eigenstate of $H - \mu n$, and $c^\dagger_{\alpha,f}$ is the electron creation operator of orbital $\alpha$ ($\alpha = 1, 2$) and spin/valley flavor $f$ ($1 \leq f \leq 4$). The average electron filling can be calculated by $\nu = \langle n \rangle - 4 = \frac{\partial \ln Z}{\partial \mu} - 4$ (note that $\langle n \rangle = 4$ at the CNP), which is valid for $|\nu| \leq 4$. Here we set $U = 23$ meV, $E_0 = 8$ meV, and $k_B T = 0.5$ meV. We then plotted the spectral weight as a function of the energy bias $\omega$ and the average filling $\nu$, and the chemical potential $\mu$ was plotted as a function of $\nu$. To simulate the remote band contribution, we assumed $\mu = \pm[\Delta + D^{-1}(|\nu| - 4)]$ for filling $|\nu| \geq 4$ (± signs for electron side and hole side, respectively), where $\Delta$ (set to 15 meV in the calculations) is the gap from the flat band to the remote band, and $D$ is the DOS of the remote band. From the definition of spectral weight, the contribution of one-particle excitations from a state with energy $\epsilon_l$ is reduced by $e^{-\beta(\epsilon_l - \epsilon_0)}$, where $\epsilon_0$ is the ground-state energy. This exponentially suppresses any spectral weight beyond energy $U$ (or $U + 2E_0$ at the CNP).

For Extended Data Fig. 3, we performed exact diagonalization (ED) of a toy Hubbard model of two nearly flat bands with 2-fold degeneracy. The model is defined on a 2 x 2 periodic triangular lattice (4 sites in total, representing the lattice of AA stacking sites) with two 2-fold degenerate orbitals (i.e. 4 orbitals in total) per site, respectively. The Hamiltonian is given by

$$H = \sum_i \left[ E_0(n_{2,i} - n_{1,i}) + \frac{U}{2} n_i (n_i - 1) \right] + t \sum_{\langle ij \rangle} \sum_{f=1,2} (c_{1,f,i}^\dagger c_{1,f,j} + c_{2,f,i}^\dagger c_{2,f,j} + h.c.)$$

where $i$ runs over sites, $\langle ij \rangle$ represents nearest neighbors, $f = 1, 2$ is the flavor index, $n_{\alpha,i} = \sum_{f=1,2} c_{\alpha,f,i}^\dagger c_{\alpha,f,i}$ is the number of electrons in orbital $\alpha$ ($\alpha = 1, 2$) at site $i$, and $n_i = n_{1,i} + n_{2,i}$ is the total number of electrons at site $i$. Extended Data Fig. 3a is the zero-temperature spectral weight on a particular site for all possible fillings. The chemical potential $\mu = dE_{gs}(N)/dN$ is determined by the ground state energy $E_{gs}(N)$ as a function of the total electron number of the lattice $N = \sum_i n_i$. We set $U = 23$ meV, $E_0 = 8$ meV, and $t = 0.6$ meV. When both flat bands are fully empty or fully occupied, and $E_F$ is in the remote band, we assume the spectral weight only shifts in bias linearly with respect to the electron filling, with slopes determined by the DOS of the remote bands.

To extract the remote band onset energies $E_{R+(-)}(v)$, we analyzed spectra $dI/dV(V_b, V_g)$ at each gate voltage $V_g$ individually. For each bias voltage $V_b$ of this spectrum, we obtain a quantity $f(V_b)$ by performing linear regressions via a least-squares method on the spectrum in two domains ($[V_b, V_b + \Delta V]$ and $[V_b - \Delta V, V_b]$, where $\Delta V$ is dependent on the gate voltage of the spectrum) and then subtracting the slopes obtained from these regressions. Local maxima of this function $f(V_b)$ occur when the slope of the spectrum discontinuously jumps at the onset of one of the remote bands. A list of these maxima and their associated gate voltages produces the data plotted in Fig. 3c, which has been averaged over a sliding window of width $\Delta V_g = 3$ V. Our mean-field model assumes a form for these onset energies $E_{R+(-)}(\nu) = \Delta_{R+(-)} + \nu U' - \mu(\nu)$. As $\nu$ increases by 1, the chemical potential increases by $U$, and the mean-field interaction term increases by $U'$. The peak-to-peak energy separation of neighboring cusps of $E_{R+(-)}(\nu)$ has a value $U - U' \approx 8$ meV, estimated from the most prominent cusps visible in Fig. 3a. Using the measured value $U = 23$ meV obtained by the extent of the cascade features, we estimate a value $U' \approx 15$ meV and solve for $\Delta\mu(\nu) = \nu U' - E_{R+(-)}(\nu)$ (which is plotted in Fig. 3d).

**Extended Data Figure Captions:**

**Extended Data Figure 1 | d$I$/d$V$ curves on an AA site. a**, Optical micrograph of MATBG/h-BN device from which data in main text Fig. 1 were acquired (Device A). The white dashed line encloses the MATBG area. **b**, STM topographic image of 1.06° MATBG ($V_b$ = -500 mV, $I$ = 10 pA). **c**, d$I$/d$V$ spectra acquired at the center of an AA site, where both flat bands are completely filled (red), the valence flat band is filled but the conduction flat band is empty (purple), and both flat bands are completely empty (blue). These data were acquired on the same AA site as that shown in main text Fig. 1c. Sequential curves are vertically offset by 20 nS for clarity. **d**, Same as **c** but for more gate voltages ($V_g$ = 50 V to -50 V). Sequential curves are vertically offset by 2.5 nS for clarity. A feature due to inelastic tunneling is sometimes seen at $V_b$ = ±60 mV but has no influence on observations closer to $E_F$ than ±60 mV[36]. The d$I$/d$V$ values for spectra with negative $\nu$ appear to be generically larger than for positive $\nu$. This asymmetry is caused in part by the constant-current feedback condition in STS measurements. As $\nu$ decreases, the flat bands are emptied and contribute less to the tunneling current at setpoint; the tip moves closer to the sample to compensate for the loss in current, amplifying the value of d$I$/d$V$ as $\nu$ decreases. We cannot however rule out the possibility that intrinsic electron-hole asymmetry also plays a role in the difference in d$I$/d$V$ between positive and negative $\nu$. **e**, -d$E_H$/d$V_g$ as a function of gate voltage (rolling average over a $\Delta V_g$ = 0.65 V window) obtained by extracting the energy $E_H$ of the peak in d$I$/d$V$ for the cascade features shown in main text Fig. 1. Upward pointing triangles identify discontinuous transitions between each cascade feature in the conduction flat band (red), around charge neutrality (purple), and in the valence flat band (blue). **f**, d$I$/d$V$($V_b$ = 0 V, $V_g$) on the AA site. Equally spaced shaded bars are drawn over dips in the zero-bias conductance, which allow us to identify each integer filling $\nu$ of the flat bands. Zero-bias conductance peaks are marked by

downward pointing triangles, which can be attributed to each of the discontinuities in **a**.

**Extended Data Figure 2 | Cascade features at other angles near the magic angle.** The cascade features that we report in this manuscript were reproduced 8 times in Device A and 9 times in Device B (all angles between 0.97° and 1.17°). Each observation can be regarded as independent because (i) the STM tip is different between each measurement and (ii) the local twist angle is different between each measurement due to large spatial inhomogeneity in each device. A small moiré between graphene and h-BN (varying around 8°) was often seen in Device A, while no graphene/h-BN moiré was seen in Device B. Here, **a-d** along with main text Fig. 1c-d show our first five observations of the cascade features in Device A. **a**, The left panel shows $dI/dV(V_b, V_g)$ measured at the center of an AA site, while the right panel shows the same quantity divided by $I/V_b$. The twist angle is 1.01° with 0.2% strain (Initial tunneling parameters: $V_b$ = -300 mV, $I$ = 0.3 nA, 551.7 Hz sinusoidal modulation of $V_{rms}$ = 1 mV). **b**, Same as **a** for a twist angle of 0.97° and 0% strain (Initial tunneling parameters: $V_b$ = -80 mV, $I$ = 1.0 nA, 4.117 kHz sinusoidal modulation of $V_{rms}$ = 0.3 mV). **c**, Same as **a** for a twist angle of 1.03° and 0.2% strain (Initial tunneling parameters: $V_b$ = -100 mV, $I$ = 0.5 nA, 4.121 kHz sinusoidal modulation of $V_{rms}$ = 1.0 mV). **d**, Same as **a** for a twist angle of 1.07° and 0.1% strain (Initial tunneling parameters: $V_b$ = -80 mV, $I$ = 1.0 nA, 4.121 kHz sinusoidal modulation of $V_{rms}$ = 1.0 mV). Twist angles are measured within ±0.01°.

**Extended Data Figure 3 | Exact diagonalization. a**, Theoretical spectral weight for different values of the filling $v$ calculated by exact diagonalization of the two-flavor, two-band Hubbard model consisting of 4 sites, interband separation $2E_0$ = 16 meV, nearest-neighbor hopping $t$ = 0.6

meV, and on-site Coulomb repulsion $U = 23$ meV. Dashed black lines highlight the cascade features, which are fewer in number than in the experimental data because the theoretical model has only 2 flavors (compared to fourfold degeneracy in MATBG). **b**, The calculated chemical potential as a function of $\nu$ shows jumps at integer fillings. The inset schematically depicts the 4 AA sites, with on-site repulsion $U$ and hopping $t$. **c**, Calculated energy difference between cascade features and the chemical potential jump at $\nu = 1$ as a function of $U$.

**Extended Data Figure 4 | Undertwisted and overtwisted bilayer graphene. a**, $dI/dV(V_b, V_g)$ measured at the center of an AA site for 0.85° (0.3% strain) twisted bilayer graphene (Device B; Initial tunneling parameters: $V_b = -40$ mV, $I = 30$ pA, 381.7 Hz sinusoidal modulation of $V_{rms} = 1.2$ mV). These data do not show the cascade features. **b**, Same as **a** but divided by $I/V_b$. **c**, $dI/dV(V_b, V_g)$ measured at the center of an AA site for 1.21° (0.2% strain) twisted bilayer graphene (Device A; Initial tunneling parameters: $V_b = -100$ mV, $I = 1.5$ nA, 4.121 kHz sinusoidal modulation of $V_{rms} = 0.5$ mV). These data do not show the cascade features (except possibly very faintly for negative $\nu$). A dip of unknown origin at $E_F$ is visible at all carrier densities. The persistence of this dip disqualifies it from being a superconducting or correlated-insulating gap. This dip is likely a Coulomb gap or a tip artifact[7]. **d**, Same as **c** but divided by $I/V_b$.

**Extended Data Figure 5 | d$I$/d$V$ on an AB site and on an AA site in a magnetic field. a**, $dI/dV$ spectra acquired at the center of an AB region in 1.05° MATBG on Device B ($V_g = 38$ V to -40 V). These data are the same as shown in main text Fig. 3a and clearly show gate-dependent oscillations in the remote bands. Sequential curves are vertically offset by 1 nS for clarity. **b**,

d$I$/d$V$ curves acquired at the center of a nearby AA site at zero magnetic field (Initial tunneling parameters: $V_b$ = -150 mV, $I$ = 0.5 nA, 4.121 kHz sinusoidal modulation of $V_{rms}$ = 1 mV). Sequential curves are vertically offset by 10 nS for clarity. **c**, d$I$/d$V$($V_b$, $V_g$) measured at the center of the same AA site as in **b** at zero magnetic field. **d**, d$I$/d$V$($V_b$, $V_g$) measured at the center of an AA site at $B_\perp$ = 9 T (Initial tunneling parameters for **c** and **d**: $V_b$ = -100 mV, $I$ = 0.5 nA, 4.121 kHz sinusoidal modulation of $V_{rms}$ = 0.6 mV). The out-of-plane magnetic field suppresses the cusps at integer $\nu$ and strongly modifies the cascade features near $E_F$. This suggests that a perpendicular field may change the localized nature of electrons in the flat bands; Landau quantization of the remote bands may also change their interactions with the flat bands.

**Extended Data Figure 6 | Tip-induced collapse of flat band.** Measurements on MATBG can strongly depend on the STM tip, so we calibrate the tip against the Shockley surface state of Cu(111). Damage to or contamination of the tip causes distortions to observed features. Copper and tungsten, incidentally, have work functions similar to graphene. **a**, d$I$/d$V$($V_b$, $V_g$) measured at the center of an AA site in Device A, with a twist angle of 1.01° and 0.2% strain (same data as Extended Data Fig. 2a). The lack of symmetry in d$I$/d$V$($V_b$, $V_g$) indicates that the tip condition is already nonoptimal. **b**, Same as **a** after lightly crashing the tip into the sample (Initial tunneling parameters: $V_b$ = -200 mV, $I$ = 0.5 nA, 551.7 Hz sinusoidal modulation of $V_{rms}$ = 1 mV). The crash further degraded the tip, causing the valence flat band to appear to cross $E_F$ without pinning. This can be caused by changes in tip-induced band bending[7] and tip-induced doping[37].

**Extended Data Figure 7 | Device fabrication. a**, Schematic showing step-by-step device fabrication procedure for Device A (Device B is similar, but without PMMA). **b**, In order to do

STM measurements, MATBG must be "face up." To achieve this, we transfer a MATBG/h-BN stack from its initial PVA handle to a secondary handle. Water injected between the two glass handles via a syringe dissolves the PVA. **c**, 500 x 500 nm$^2$ STM topograph of Device B showing a moiré pattern and physical corrugation in the underlying substrate ($V_b$ = -400 mV, $I$ = 100 pA). **d**, Zoomed-in image of the center of the area in **c**. Each device has regions that are magic angle (which always show cascade features) and regions that are not magic angle (which do not show cascade features). Regions of a device that are magic angle are typically so over at least a square micrometer.

**FIGURE 1**

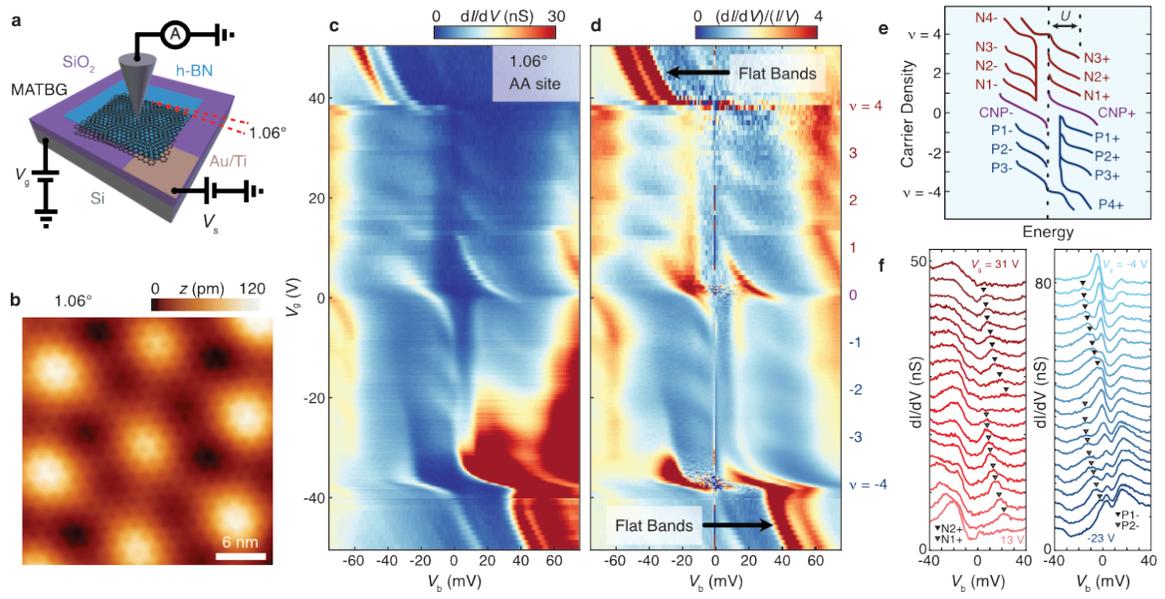

**FIGURE 2**

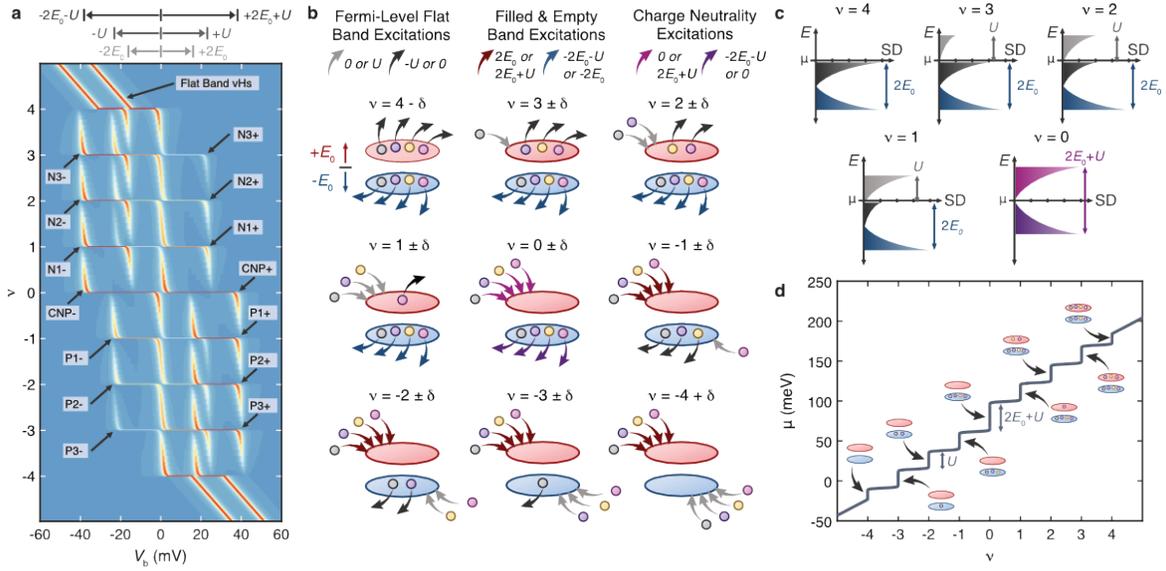

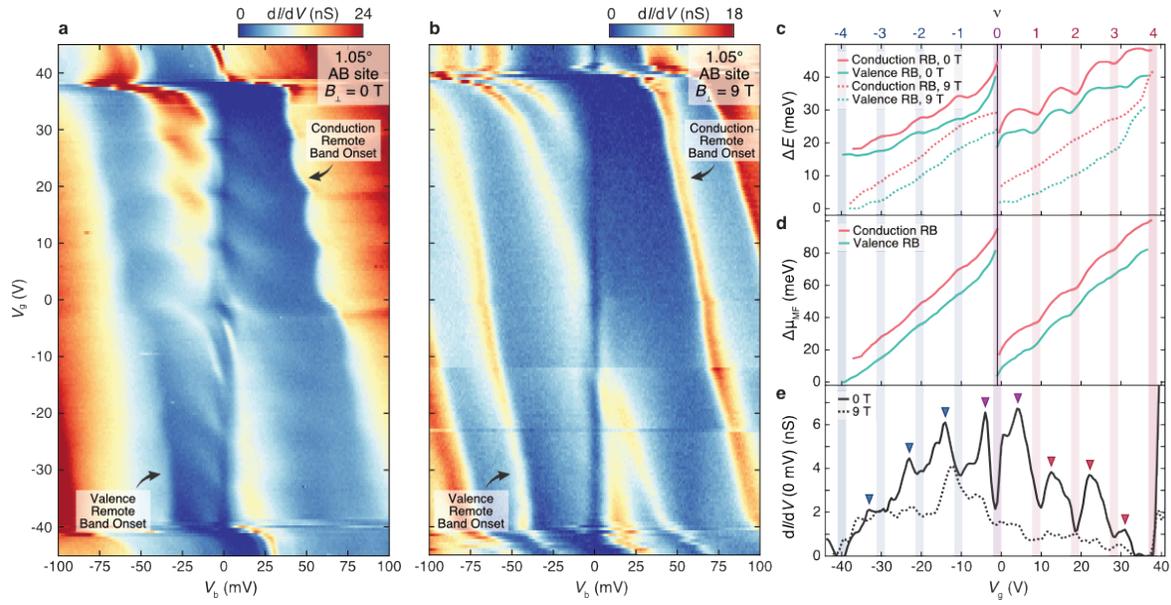

# EXTENDED DATA FIGURE 1

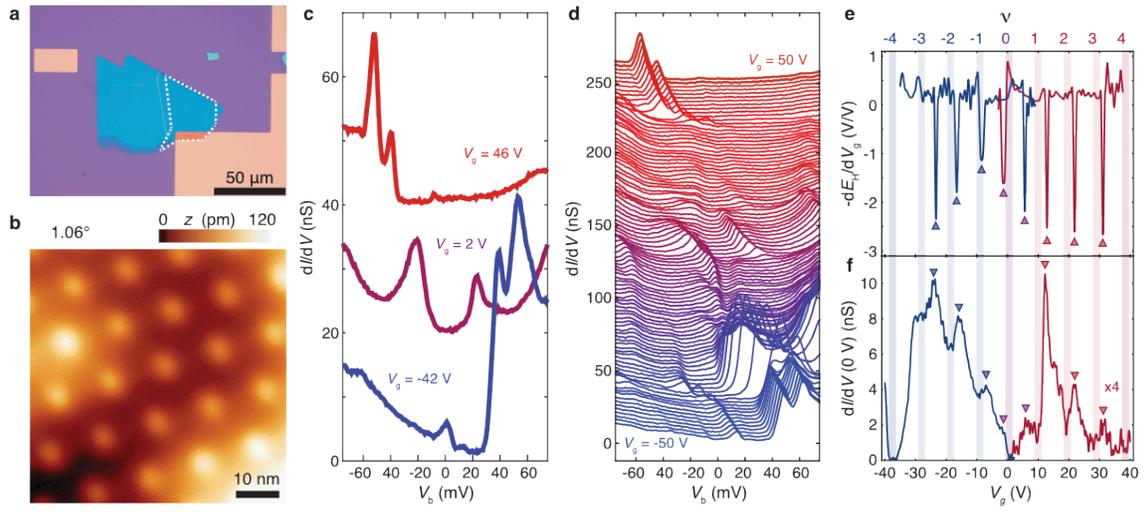

# EXTENDED DATA FIGURE 2

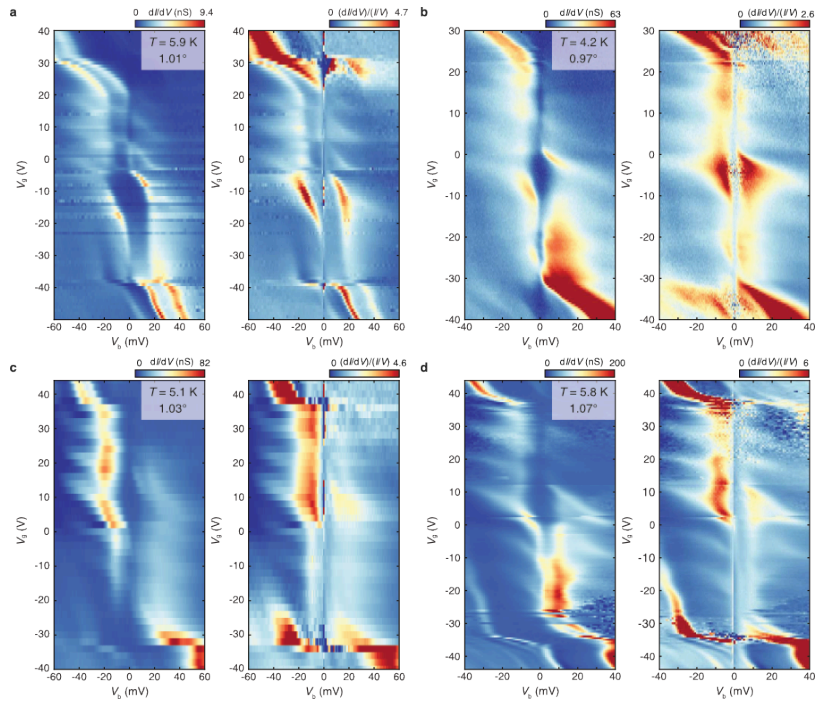

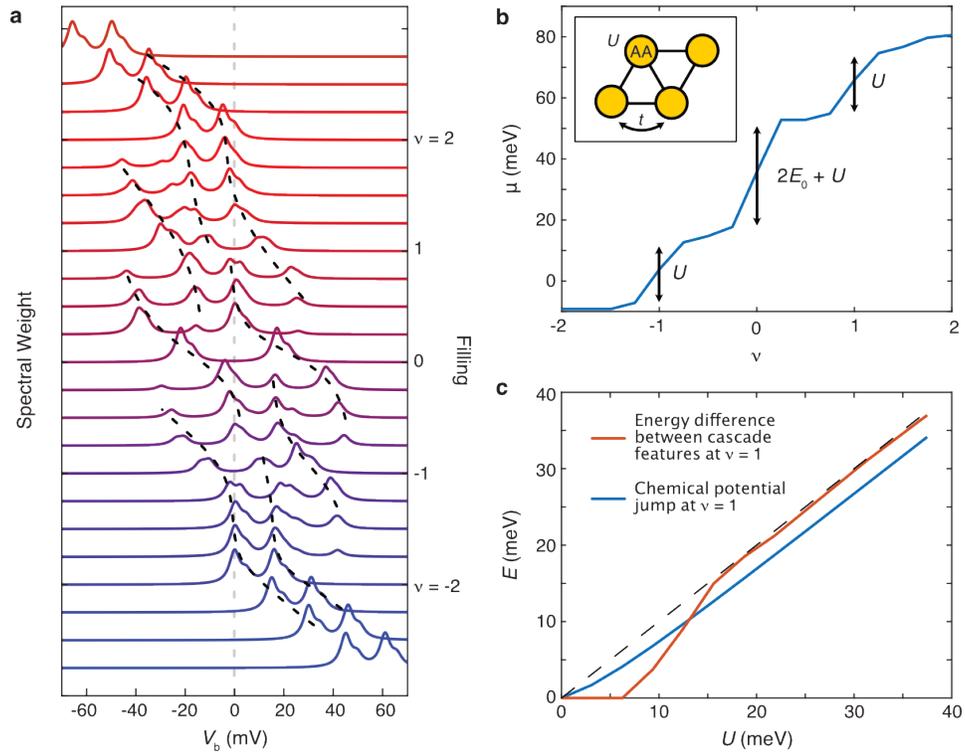

# EXTENDED DATA FIGURE 4

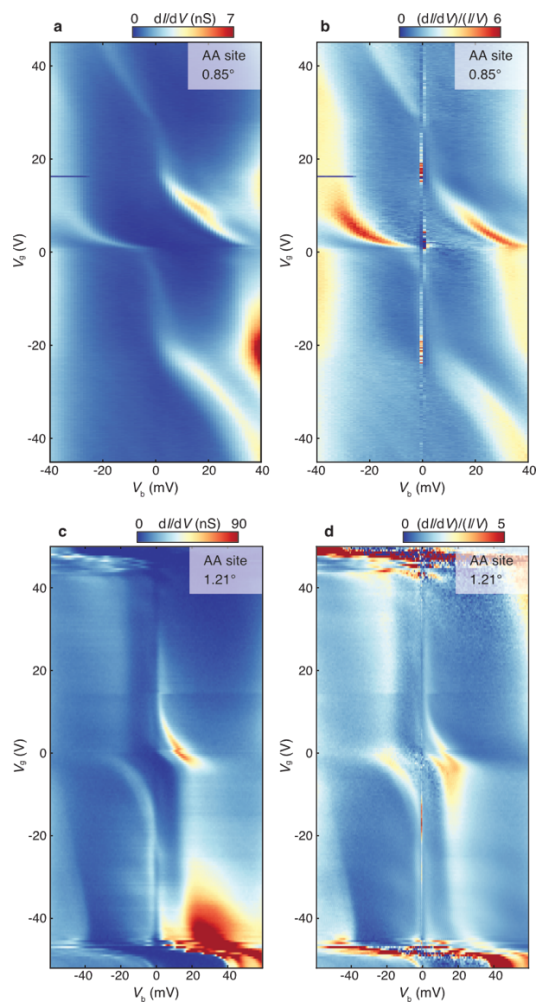

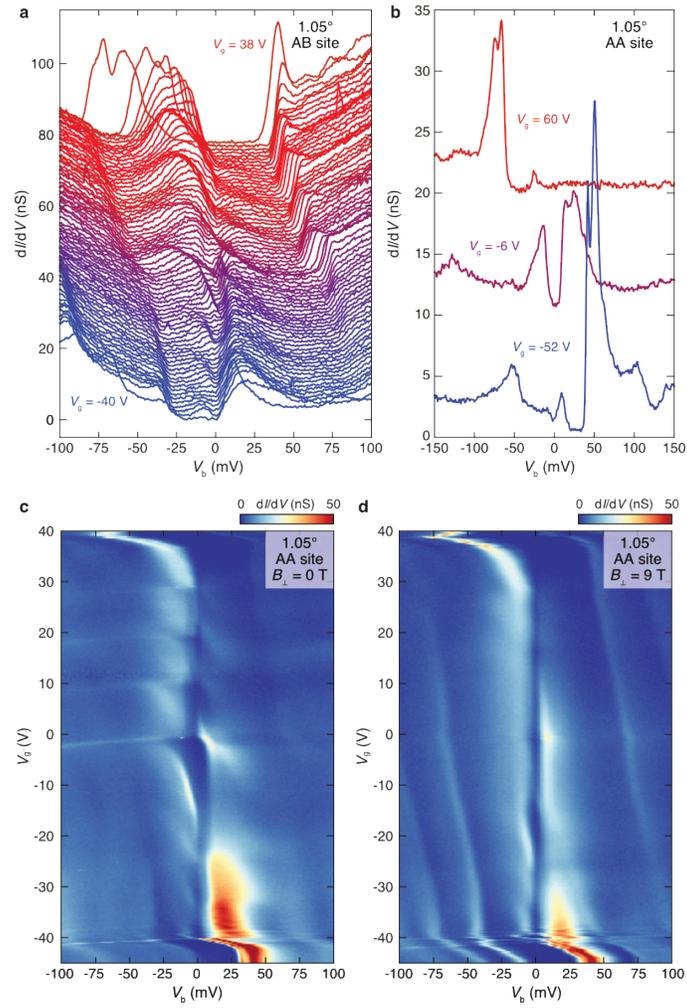

# EXTENDED DATA FIGURE 6

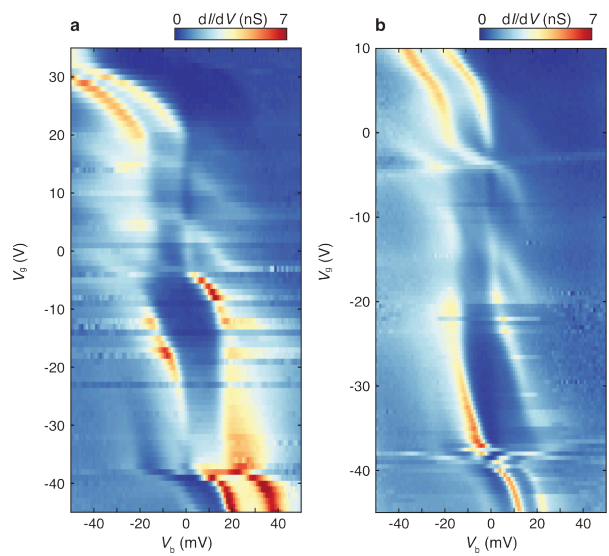

# EXTENDED DATA FIGURE 7

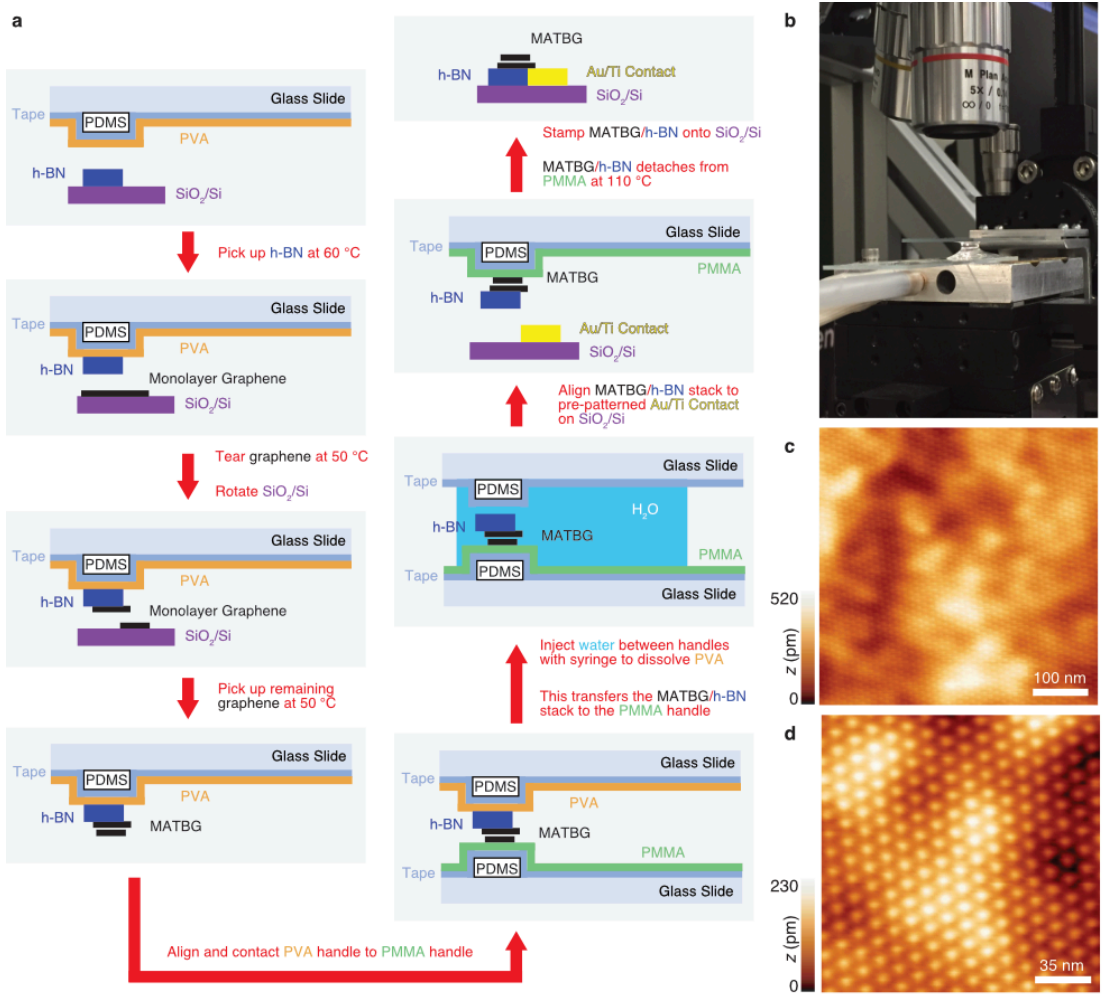